\documentclass[final,5p,times,twocolumn,sort&compress]{elsarticle}

\usepackage{graphicx}

\usepackage{amsmath}
\usepackage{amssymb}
\usepackage{textcomp}

\usepackage{wasysym}

\usepackage{lineno}

\usepackage{dblfloatfix}

\usepackage{verbatim}  
\usepackage{epstopdf}
\usepackage{hyperref}  
\usepackage{breakurl}
\hypersetup{breaklinks = true, colorlinks = true, citecolor = blue, linkcolor = blue, urlcolor = blue}
\journal{Nucl. Instr. and Meth. A}

\begin{document}

\begin{frontmatter}

\title{A short-orbit spectrometer for low-energy pion detection in electroproduction experiments at MAMI}

\author[kph]{D. Baumann}
\author[kph]{M. Ding}
\author[pmf]{I. Fri\v{s}\v{c}i\'{c}\corref{cor1}\fnref{fn1}}
\author[kph]{R. B\"{o}hm}
\author[pmf]{D. Bosnar}
\author[kph]{M. O. Distler}
\author[kph]{H. Merkel}
\author[kph]{U. M\"{u}ller}
\author[kph]{Th. Walcher}
\author[kph]{M. Wendel}

\cortext[cor1]{Corresponding author, Email: friscic@mit.edu}
\fntext[fn1]{Present address: MIT-LNS, Cambridge MA, 02139, USA}
\fntext[fn2]{This paper comprises parts of the doctoral theses of D. Baumann, M. Ding and I. Fri\v{s}\v{c}i\'{c}.}

\address[kph]{Institut f\"{u}r Kernphysik, Johannes Gutenberg-Universit\"{a}t, 55099 Mainz, Germany}
\address[pmf]{Department of Physics, Faculty of Science, University of Zagreb, Bijeni\v{c}ka c. 32, 10000 Zagreb, Croatia}

\begin{abstract}
A new Short-Orbit Spectrometer (SOS) has been constructed and installed within the experimental 
facility of the A1 collaboration at Mainz Microtron (MAMI), with the goal to detect low-energy 
pions. It is equipped with a Browne-Buechner magnet and a detector system consisting of two 
helium-ethane based drift chambers and a scintillator telescope made of five layers. The detector 
system allows detection of pions in the momentum range of 50 -- 147 MeV/c, which corresponds 
to 8.7 -- 63 MeV kinetic energy. The spectrometer can be placed at a distance range of 54 -- 66 cm 
from the target center. Two collimators are available for the measurements, one having 1.8 msr 
aperture and the other having 7 msr aperture. The Short-Orbit Spectrometer has been successfully 
calibrated and used in coincidence measurements together with the standard magnetic spectrometers 
of the A1 collaboration.

\end{abstract}


\end{frontmatter}

\section{Introduction} \label{introduction}

The $\mathrm{p(e,e'\pi^+)n}$ reaction, measured with high precision close to threshold, 
provides a great tool for studying the axial and the pseudoscalar form factor \cite
{amaldi_79,drechsel_92}. One of the experimental setups capable performing such experiments 
is the high resolution spectrometer setup of the A1 collaboration \cite{blomqvist_98} using 
the electron accelerator Mainz Microtron (MAMI) \cite{kaiser_08}.

The A1 spectrometer setup consists of two large acceptance (28 msr) 
spectrometers (A and C) having a quadrupole-sextupole-dipole-dipole 
magnet configuration and a third spectrometer (B), which is based on 
one clamshell dipole. This spectrometer is slimmer and can reach 
scattering angles down to 7$^\circ$, but this comes at the cost of 
smaller acceptance (5.6 msr). Each spectrometer is equipped with a 
particle tracking system based on 4 vertical drift chamber layers 
and a particle identification system based on scintillating paddles 
and \v{C}erenkov detectors. If needed, these can be replaced with 
other detector configurations. The spectrometers can be rotated around 
the target chamber in the center and they can be operated in a single, 
double, and triple coincidence mode. Looking along the beam direction, 
spectrometers A and C are placed on the left and right side of the beam 
pipe, respectively. Spectrometer B can be moved to either side of the 
beam pipe; in standard configuration it is placed on the right side. 
All three spectrometers have a relative momentum resolution $\le$ 10$^{-4}$ 
and are used in different electron scattering experiments for detection 
of charged particles having momenta up to 735 MeV/c in A, 870 MeV/c in B 
and 551 MeV/c in C.  For unstable particles the use of spectrometers 
is limited to high momenta, which allow the particles to reach the 
detector system before they decay.

The pion is an unstable particle having a lifetime of 26.033 ns, which dominantly 
decays into a muon and a muon neutrino \cite{beringer_12}. Close to the reaction threshold 
it will have low kinetic energy. Since the distance from target to standard detector system 
inside the spectrometers of the A1 collaboration is of the order of 10 m, most of the low energy 
pions will decay before they reach the detectors. The measurement time needed for collection of 
statistically relevant data would be prolonged. Moreover, a fraction of muons will be created 
close to the direction of pions and will be also detected. Such muons cannot be distinguished 
from the pions and they are a major source of systematic error. The influence of these two 
problems can be reduced simultaneously by decreasing the distance between target and 
detector system. Significantly more pions will now survive to the detector system for the 
same kinetic energy, but the data still has to be corrected for pion decay. At the 
same time, muon contamination of the data will be reduced, so that it is feasible to determine 
it from a simulation and correct for it.

Therefore, a new short orbit spectrometer (SOS) was constructed specifically for detection 
of low energy pions. The $\mathrm{^{12}C(e,e'p)^{11}B}$ reaction was used for calibration 
of the SOS. As the first physics experiment, the $\mathrm{p(e,e'\pi^+)n}$ reaction was evaluated, 
which at the same time allowed to test the performance of the SOS in detection of low energy pions.

Section \ref{layout} of this paper describes the magnet, detector system and electronics 
of the SOS. In section \ref{calibration} the measurement used for 
calibration of the SOS is discussed. Section \ref{electroproduction} deals with a measurement of the  
$\mathrm{p(e,e'\pi^+)n}$ reaction. A summary and a conclusion 
are done in section \ref{sec:sum_conc}.

\section{Short-orbit spectrometer outline} \label{layout}

\subsection{Construction objectives}

   The objectives of the spectrometer's construction arise from the
   following experimental considerations:
\begin{enumerate}

   \item  In order to minimise the above mentioned loss due to pion decay and at
   the same time reduce muonic background, the spectrometer has a
   flight path of only 1.6 m from target to detector system.

   \item Good missing mass resolution, e.g. for the neutron in the
    $\mathrm{p(e,e'\pi^+)n}$ reaction, is desirable, because it will help to identify
   the reaction's final state, as well as to reduce background noise. For this, the
   SOS must be able to detect low energy pions with good spatial and
   momentum resolution. This requires focusing at least in the
   dispersive plane of the spectrometer magnet as well as a position
   sensitive detector. The reaction vertex in the target can be
   determined independently, using information of the beam position and
   data from the electron spectrometer.

    \item The detector needs to have a small effective mass thickness to
   minimise multiple scattering and energy loss of particles down to
   low energies. Therefore, a drift chamber with a helium based counting gas was
   chosen as tracking detector.

   \item For further reduction of charged particle background, the
   detector should allow particle identification, i.e. discrimination
   of pions from protons, electrons or positrons. This is achieved
   by a scintillator range telescope. In addition, time of flight is
   measured. But in the pion arm this is of limited use due to the short
   path length. Use of an aerogel or plastic \v{C}erenkov counter was
   rejected because it would offer no advantage over the range telescope
   in the intended momentum range.

   \item  Solid angle acceptance should be reasonably high. However,
   short path length was deemed more important for the considered
   experiments. Therefore, the spectrometer does not feature
   additional focusing by a quadrupole, which limits its solid angle
   to at most 7 msr. Furthermore, measurements with an extended
   target, like a liquid hydrogen target cell with 2 cm diameter,
   should be possible.

   \item The spectrometer should be movable over a wide angle range in
   order to maximise the available kinematics, e.g. the variation
   range of the polarisation parameter $\epsilon$ \cite{drechsel_92}. 
   This implies small relative angles to the beamline, as well as to the other
   spectrometers.

\end{enumerate}

\begin{figure}[!h]
\begin{center}
\includegraphics[height=310pt, width=180pt]{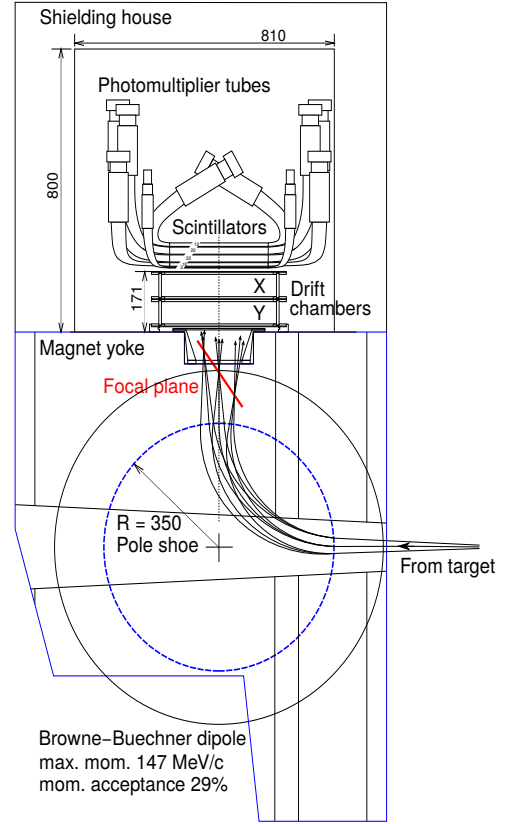} \centering 
\caption{(colour online) Schematic of SOS and the detector system. All measures are 
in millimeters. The target is positioned approx. 1.1 m to the right and aligned with the central 
cross.}
\label{Fig1}
\end{center}
\end{figure}

\subsection{Magnet}

\begin{figure}[!h]
\begin{center}
\includegraphics[height=345pt, width=140pt]{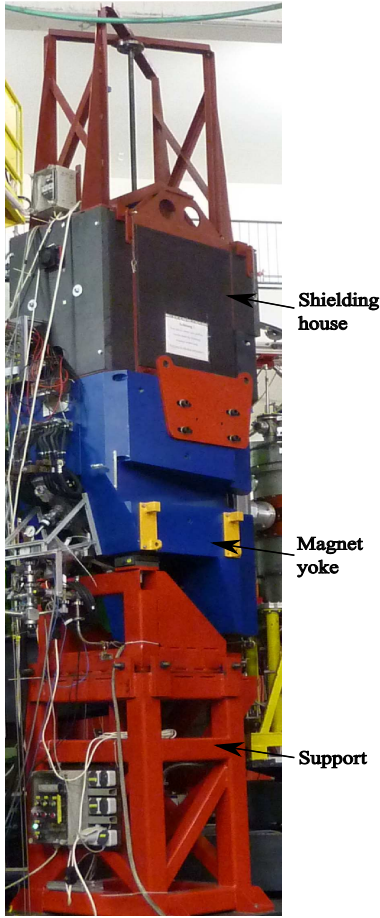} 
\caption{(colour online) Photo of the SOS. The stand of the SOS is mounted on the support 
of spectrometer B. It carries the magnet yoke and it allows fine alignment 
of the yoke with respect to the target center. Above the yoke is the shielding 
house containing the detector system, which can be accessed for maintenance 
by lifting the sides of the shielding house.}
\label{Fig2}
\end{center}
\end{figure}

The SOS, see Fig. \ref{Fig1} and \ref{Fig2}, comprises a Browne-Buechner type dipole magnet 
\cite{magnet_BB} having circular pole pieces with a radius of $350$ mm.  
The magnet from the previous Mainz pion spectrometer was available and 
could be used with some modifications. See ref. \cite{schmitt_1985} for 
a detailed description of the magnet.

The magnet has three focal planes. Two of them correspond to $\pm$90$^\circ$ 
deflection of the reference trajectory and can be used for momenta less than 
150 MeV/c. They allow detection of positive and negative charged particles 
simultaneously. The third focal plane is located in the rear lower part of the SOS. 
It can be used for particles having momenta in range of 150 -- 300 MeV/c. At the 
present time, only the upper focal plane is equipped with a detector system. 
The spectrometer was designed to accept particles with momenta up to 147 MeV/c 
for a magnetic field of 1.4 T. The momentum acceptance ranges from $-13\%$ to 
$16\%$. The SOS is mounted on the support of spectrometer B, which allows 
adjustment of the spectrometer angle with a 0.01$^{\circ}$ resolution. During 
the measurement with the SOS, spectrometer B cannot be used.

The shape of the magnet yoke was modified by milling in order to allow for a 
smaller angle relative to the exit beamline, thereby maximizing the $\epsilon$ 
range. Also the smallest possible angle relative to the other spectrometers 
was diminished. Before the milling, the influence of the yoke's shape on the 
homogeneity of the magnetic field was investigated via computer simulation. 
After the reassembly of the magnet the homogeneity of the magnetic field was 
measured for several strengths of the field. No significant changes were 
observed before and after the modification.
For a distance of 66 cm between SOS and target center the modification 
of the yoke's shape allows the minimum relative angle between SOS and spectrometers 
A and C of $55^{\circ}$. The minimum forward angle is $15.4^{\circ}$. At a 
distance of $54$ cm the minimum forward angle is 22$^{\circ}$. 

It is possible to directly connect the vacuum system of the magnet and the 
scattering chamber. In this way the need for two additional foils, as well as the air 
gap between the two foils can be avoided. The accessible angles are 
limited then, due to several discrete openings of the scattering chamber. A change of
angular position requires uncoupling of the SOS from the scattering 
chamber. This includes a breach in the vacuum system.
In this work due to running time and angle limitations, the scattering 
chamber and the magnet vacuum were not directly connected.

The vacuum system of the magnet ends with a tapered flange, sealed with a 
50 $\mu$m thick polyimide foil ("Kapton"). The bending of the Kapton foil 
caused by the vacuum stress leads to formation of an air pocket in the 
created free space. This volume is flushed with helium gas for further 
reduction of multiple scattering. The upper cover, just below the 
drift chambers, is made out of 6 $\mu$m aramid foil having only 
small helium permeability.

\begin{table*}[!ht]
\caption{Contributions to effective thickness of SOS tracking detector, 
$x$ is thickness of a layer and $X_0$ is radiation length of the employed material.} 
\begin {center}
\begin {tabular} {l|c|c|c|l}
\hline
Material        &  $x$                  & $X_0$ (cm) & $x/(10^{-6} X_0)$& Comment \\ \hline
He + C$_2$H$_6$ & 152 mm                & 63900      & 238              & Counting gas \\
PET             & (4+4+12) $\mu$m       & 28.7       & 69.7             & Chamber foils (CF) \\
Al              & 2$\cdot$(10+10+40) nm & 8.9        & 1.35             & Coating on the CF \\
Al              &367 nm                 & 8.9        & 4.1              & Potential wires  (averaged)\\ \hline
Drift chambers  &                       &            & 313              & Sum for the chambers \\ \hline
Polyimid        & 50 $\mu$m             & 28.6       & 175              & Vacuum exit \\
Aramid          & 6 $\mu$m              & 28.6       & 21.0             & He-flange foil \\
Helium          & 20 mm                 & 568000     & 3.5              & Gap below the chamber \\ \hline
Rest            &                       &            & 277              & Everything outside the chamber \\ \hline
Altogether      &                       &            & 590              & From target to scintillator \\ \hline
\end {tabular} 
\end {center}
\label{Tab1}
\end{table*}

\subsection {Collimators}

Altogether, the SOS is equipped with three different collimators. Two are meant 
to be used in experiments and the third one is a "sieve" collimator for
determination of transfer matrices. The collimators were developed for 
the measurements with an extended target. The size of the collimator's 
aperture was optimized with respect to the distance from target, so 
that the accepted solid angle was maximized and the amount of internal 
scattered particles was minimized. However, since a Browne-Buechner magnet 
is only single-focusing in the dispersive plane, the maximum achievable 
solid angle is limited.

One collimator was developed for experiments in which the SOS will be placed 
at $66$ cm from the target center, while a target cell with diameter of 
$2$ cm will be used. To ensure that no particles from the target cell will 
hit the pole shoes, the solid angles of this collimator was limited to 
1.8 msr \cite{baumann_phd}.

For a certain angle range between the limiting values (described above) the 
SOS can be also mounted closer to the target. On one hand, a larger solid 
angle provides a reduction of measurement time. On the other hand, due to 
shorter orbits a smaller number of pions will decay into muons. The distance 
of 54 cm was found to be an optimal value. But for this distance it was not 
possible to design a collimator, which ensures that no particles from the 
target hit the magnet poles. Due to this fact the collimator aperture of 
7 msr is reduced to approximately 4 msr \cite{baumann_phd}. Particles 
scattered on the magnet poles have to be taken into account during the 
analysis.

The third collimator is a sieve collimator with an irregular array of holes. 
It is used for a (inverse) transfer matrix determination.

All collimators are made of the same tungsten-copper alloy ("Densimet 18" by 
Plansee). The thickness of the sieve collimator and the other two collimators 
is 5 mm and 45 mm, respectively.

\subsection {Drift chambers}

\begin{figure}[!ht]
\begin{center}
\includegraphics[height=180pt, width=240pt]{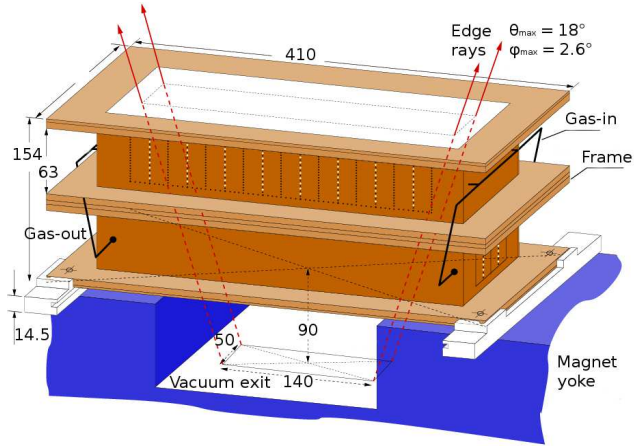}
\caption{(colour online) The SOS drift chambers in a build-in position. The bottom chamber is used 
for reconstruction of non-dispersive coordinates, and the upper chamber for dispersive 
coordinates. Positions of the signal wires are marked with white points. Edge rays were 
determined with help of RAYTRACE simulation.} 
\label{Fig3}
\end{center}
\end{figure}

Since a part of the magnet's focal plane is situated inside the influence of fringe 
fields, the tracking detector has to be placed outside the focal plane. 
Therefore, the tracking detector was realized by using two volume type drift 
chambers, see Fig. \ref{Fig3}. 

The effective thickness was kept small by reducing the number of foils to 
three: two outer and one shared. The shared foil was grounded in order to 
separate the electric field configurations of upper and lower chamber. 
Furthermore, due to its larger radiation length than argon, helium was chosen 
as a counting gas and ethane as a quencher. During the measurement the chambers 
are operated with 1:1 gas ratio, ensuring in this way almost constant drift 
velocity for a wide range around the working voltage \cite{ding_phd}. Aging 
of the wires is inhibited by enriching the gas mixture with ethanol before 
it enters the chambers. Thickness and radiation length of the employed  
materials are given in Table \ref{Tab1}.

The chamber frame is made of epoxy reinforced with fibreglass
(Stesalit AG, CH-4234 Zullwil). The wires are tensioned and then held
in place by crimping them in feedthrough ferrules (Ferrini S.A., CH-6596
Gordola). These have an inner bore of 80 $\mu$m or 150 $\mu$m in order to
center the signal and potential wires, respectively. The ferrules also
provide electrical contact for the wires. Additional details are listed 
in Table \ref{Tab2}.

\begin{figure}[!ht]
\begin{center}
\includegraphics[scale=0.26]{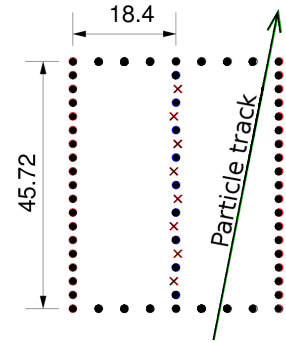}
\caption{(colour online) Lateral profile of a single SOS drift cell. Circles represent potential 
wires and crosses represent signal wires, signal wires, which are staggered left and right by 100 $\mu$m. In order to make the signal wire staggering visible on this figure, it
is exaggerated.} 
\label{Fig4}
\end{center}
\end{figure}

Fig. \ref{Fig4} exemplary shows a single SOS drift cell. Each drift cell is made 
of eight vertically arranged signal wires, which are alternating shifted by 100 
$\mu$m with respect to the center of the cell. In the trajectory reconstruction 
procedure the shifts are taken into account to determine at which side the particle 
has passed the signal wire -- so-called left-right decision. The signal wires are 
grounded by the preamplifier. The potential wires (cathode wires) at the 
edges of the drift cells are set to a high negative voltage, typically $-$5400 V. 
In order to obtain drift cells with well defined borders and to mutually suppress 
the cross talk between neighboring drift cells, potential wires closest 
to the signal wires are set to an intermediate voltage by a voltage degrader. 
The best averaged single wire efficiency of 95\% was achieved, when the ratio of 
the cathode wire and intermediate voltage was 4 : 1 \cite{ding_dipl}. In order to 
assure that the boundary drift cells also have a proper geometry of the electric 
field, only potential wires at the edges of the signal wires vertical arrangement 
are grounded.

\begin{table}[!ht]
\caption{Properties of the SOS drift chambers.} 
\begin {center}
\begin {tabular} {l|r}
\hline
Outer dimensions & 410 x 210 x 156 mm$^3$ \\
Active surface & 294.4 x 73.6 mm$^2$ \\
Wire drift cell length x width & 18.4 x 5.08 mm$^2$ \\
Signal wires horizontal offset & $\pm$100 $\mathrm{\mu}$m \\
\hline
Number of signal wires  & 64/16 (X-/Y-chamber)\\
Number of potential wires & 355/103 (X-/Y-chamber)\\
\hline
\multicolumn{2}{l}{Signal wires: $\quad\;\:$ $\diameter$15 $\mathrm{\mu}$m Au-coated tungsten-rhenium } \\
\multicolumn{2}{l}{Potential wires:  $\;\:$ $\diameter$80 $\mathrm{\mu}$m Ag-coated aluminium}  \\
\multicolumn{2}{l}{Gas:  $\quad\quad\quad\quad\quad\;$50\% helium + 50\% ethan} \\
\hline 
\end {tabular} 
\end {center}
\label{Tab2}
\end {table}

The Y-chamber with two drift cells is the first chamber above the magnet. It measures 
the non-dispersive coordinate and angle  $(y_{ch}, \phi_{ch})$. The second chamber 
consisting of eight cells is the X-chamber. The wires of the X-chamber are perpendicular 
to the wires of the Y-chamber and, therefore, it measures the dispersive coordinate 
and angle $(x_{ch}, \theta_{ch})$. 
Particles enter the X-chamber after the Y-chamber and due to multiple 
scattering, resolution of the X-chamber will always be lower relative to the Y-chamber. 
The priority to Y-chamber was given to provide better discrimination of particles 
which may be scattered from the magnet edges and to allow better estimation of 
particle energy loss inside a target.

The individual signal wires are connected to preamplifier and
discriminator cards (LeCroy 2735 DC) which are mounted to the chamber's
frame. The signals are then digitized by TDCs (TDC2001) with 2 $\mu$s
measuring range and 250 ps time resolution.

\subsection {Scintillator range telescope}

\begin{figure}[!ht]
\begin{center}
\includegraphics[scale=0.22]{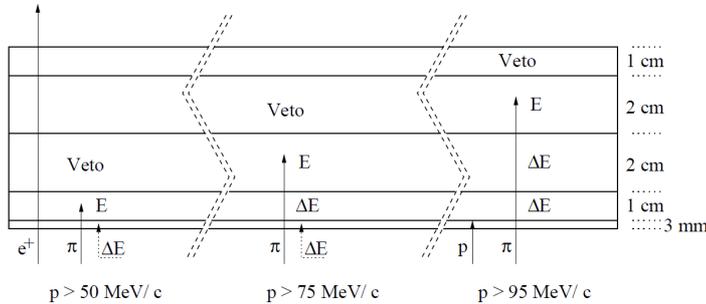}
\caption{Illustration of the particle identification in the scintillator range 
telescope. Protons are stopped in the bottom scintillator, positrons pass through 
all layers, pions are stopped in a certain layer depending one their momentum. 
In order to enter the second scintillator the pions need to have momenta larger 
than $50$ MeV/c and so on \cite{baumann_phd}.}
\label{Fig5}
\end{center}
\end{figure}

The SOS range telescope has two tasks. First, it gives a timing signal for 
the trigger and common stop for the drift chambers. Second, it allows 
particle identification by range information and particle energy 
deposition in a certain scintillator layer, see Fig. \ref{Fig5}. The range 
telescope is composed of five scintillator layers. Scintillators are made of 
Bicron BC $408$ type plastic. Each scintillator has the same width of $80$ mm 
and length of $300$ mm, but they differ in thickness. From bottom to top,
the scintillator's thickness are $3$, $10$, $20$, $20$ and $10$ mm,
respectively, see Fig. \ref{Fig5}. The scintillators are equipped with PMTs 
at both lateral ends. Read-out is performed in coincidence for background suppression. 
The thinnest scintillator is read-out by Philips XP $2910$ PMTs and all other scintillators 
by Philips XP $2262$B PMTs. 

So-called "twisted-strip" Plexiglas light guides are used to connect PMTs and 
scintillators. This type of light guide was produced by first bending the 
individual strips to the desired radius and then gluing the strips together. 
Individual strips have to be of same length, to ensure simultaneous arrival 
of the light to PMT. However, due to limitations by critical bending radii, 
as well as the compact build of the shielding house, same length for all strips 
was not feasible. Nonetheless, the length difference of the strips 
was still acceptable and was less than 2 cm. The optical coupling between 
scintillator and light guide was realized by using optical cement 
Bicron BC-600. For optical coupling between the light guide and PMT, 
two-component silicon rubber (Wacker Elastosil RT 601) was used. This glue 
allows easy detachment of a PMT if it needs to be replaced. The PMTs are placed 
on an aluminium fixture and shielded against magnetic fields using $\mu$-metal.

\subsection{Shielding house}
The detectors are located inside a shielding house which consists of an outer 
5 cm layer of borated polyethylene to absorb neutrons and an inner 10 cm layer 
of lead to shield electromagnetic background. This specific shielding configuration 
was \linebreak chosen in order to maximise the available space inside the shielding house, 
for better accommodation and easier mounting of the detector system, while still 
retaining the best possible shielding effect. Therefore, an originally planned
5 cm layer of normal, unborated polyethylene was omitted in the final design \cite{baumann_dipl}.

Since the drift chambers are not leak-proof, some amount 
of helium could diffuse into PMTs and damage them. To prevent this, the shielding 
house is divided with plastic plates in an upper (scintillator) and lower 
(drift chamber) part. Additionally, the chamber part is kept under-pressurized 
by a ventilator at the end of an exhaust tube, which connects the chamber part 
and the rear wall of the shielding house.

\subsection {Trigger and data acquisition}

In order to achieve high event rates, the SOS is equipped with
independent electronics and a trigger logic system. Each of the two
PMTs of the five scintillators provides an analog signal which is then
split, one half being sent to an ADC (LeCroy 2249A) for charge
digitization, the other half being converted by a discriminator into a
logical signal.

A minimal condition for generation of a trigger signal requires that
both PMTs of a single scintillator layer produce signals larger than a
specified threshold. This minimal trigger condition can be extended by
requesting different logical conditions between trigger signals
of individual layers, for example, by requiring generation of
trigger signals in two or more scintillators, or by using one or more
scintillator layers as veto detectors. This is achieved by using a
programmable lookup unit (LeCroy PLU 4508) which also allows an
on-line change of trigger conditions. An additional input of the PLU
is used to inhibit triggers during data readout.

\section{Calibration of SOS with quasielastic proton knockout from $^{12}$C }  \label{calibration}

\begin{figure}[!hb]
\begin{center}
\includegraphics[scale=0.22]{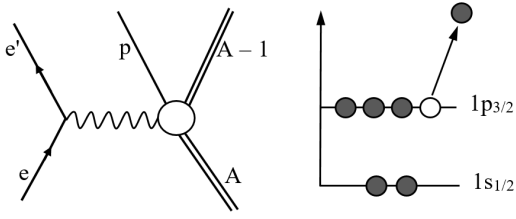} 
\caption{On the left: illustration of the quasielastic electron scattering on a 
nucleus having a mass number A  with a proton knockout. On the right: in the case 
of $^{12}$C, knowing that the remaining $^{11}$B is in the ground state ensures 
that the proton was knocked out of the p-shell.} 
\label{Fig6}
\end{center}
\end{figure}

Usually, calibration of a magnetic spectrometer is performed using the electrons from 
the elastic scattering of a selected nucleus, since in this case the momentum of a 
scattered electron is only dependent on the scattering angle. 
Because of the maximum momentum accepted by the SOS (147 MeV/c), the elastic scattered 
electron produced by the minimum electron beam energy of 180 MeV, provided by MAMI, 
does not comply with SOS requirements. Hence, another reaction has to 
be used for the calibration of the SOS. The choice for a suitable reaction was guided 
by the fact that the scattered electron has to lose enough momentum in order to comply 
with the SOS's momentum acceptance. The quasielastic electron scattering from a
$^{12}$C nucleus with proton knockout meets these demands: protons 
inside a $^{12}$C nucleus are not static, they move (Fermi-motion) having a momentum 
distribution with a maximum at 100 MeV/c \cite{frullani_84, blomqvist_95}. If the remaining 
$^{11}$B nucleus is determined to be in the ground state, it is certain that the 
outgoing proton, which is detected in spectrometer A, originates from the $1$p$_{3/2}$ 
level of the $^{12}$C nucleus, Fig. \ref{Fig6}. 

Therefore, the SOS calibration measurement has been performed using a $^{12}$C 
foil with area density of 30 mg/cm$^2$ as target. Table \ref{Tab3} summarizes 
the kinematical settings used for the calibration of the SOS.
For setting 1 from table \ref{Tab3}, the energy transfer is $\omega =$ 55 MeV, 
which is in the case of the quasielastic proton knockout from $^{12}$C divided between 
the separation energy $E_{sep} = (m_p + m_{^{11}B} - m_{^{12}C})c^2$, excitation energy 
of the residual $^{11}$B nucleus $E_{x}$, kinetic energies of the proton $E_p$ and the 
residual $^{11}$B nucleus $E_{^{11}B}$:
\begin{equation} \label{eq:eq1}
\omega = E - E' = E_{sep} + E_{x} + E_p + E_{^{11}B}
\end{equation} 
From equation (\ref{eq:eq1}) and the measured proton momentum it is now possible to determine 
the momentum of the scattered electron and perform calibration of the SOS. Four-momentum 
based calculation of recoil energy leads to an exact expression for the energy of the 
scattered electron \cite{ding_phd}. The idea is to define a four-momentum $\kappa = p_e + p_{^{12}C} -p_p$, 
which can be easily obtained by the measured quantities. Using the momentum conservation 
law one obtains:
\begin{equation} \label{eq:eq2}
p_{^{11}B} = \kappa - p'_e
\end{equation} 
Taking the square of equation (\ref{eq:eq2}) and neglecting electron mass one has:
\begin{equation} \label{eq:eq3}
E'_{calc}= \frac{\kappa^2-m^2_{^{11}B}} {2\kappa \cdot \hat{\eta}}
\end{equation} 
where the spatial part of the unit vector $\hat{\eta} = (1,sin\theta_e,0,cos\theta_e)$ points 
in the direction of the scattered electron. Equation (\ref{eq:eq3}) depends on the electron 
scattering angle $\theta_e$. Using the sieve collimator the variation of $\theta_e$ can be 
kept small enough to use equation (\ref{eq:eq3}) for the calibration of the SOS.

\begin{table}
\caption{Central kinematical parameters for the calibration of the SOS: $E$ is the beam energy, 
$p_{A}$ and $\theta_{A}$ are central momentum and angle of spectrometer A, $p_{SOS}$ 
and $\theta_{SOS}$ are central momentum and angle of SOS, respectively.} 
\begin{tabular}{cccccc} 
Setting      & $E$    & $p_{A}$   & $\theta_{A}$  &  $p_{SOS}$  & $\theta_{SOS}$  \\
            & (MeV)  & (MeV/c) & ($^{\circ}$)  &    (MeV/c)      &  ($^{\circ}$) \\ \hline 
1   & 180 & 265 &    43.0  &    125        & 60.0 \\ 
2   & 180 & 265 &    43.0  &    115        & 60.0 \\ 
3   & 180 & 290 &    43.0  &    115        & 60.0 \\ 
\end{tabular} 
\label{Tab3} 
\end{table}

The scattered electron in the SOS and the knockout proton in spectrometer A were detected in coincidence 
between those two spectrometers. The trigger in spectrometer A was formed by requiring a signal 
in both scintillator planes ($\Delta$E and ToF), but the timing information was defined only by the 
ToF plane. Coincidence time was corrected for the path length of the particle only in spectrometer 
A. After this correction, a coincidence peak with a time resolution of 1.98 ns FWHM
was obtained, which was sufficient to determine the transfer matrices for the SOS. A cut of
$-$ 2 ns $\le T_{A-SOS} \le$ 2 ns was used to select true coincidence events. 

The background of minimum ionizing particles was suppressed by imposing an appropriate cut on the
deposited energy in the $\Delta$E scintillator plane. For further reduction of background additional 
cuts were applied on the angular acceptance of spectrometer A ($|\theta_0| \le 5.1^{\circ}$ and 
$|\phi_0| \le 0.6^{\circ}$).

Using the measured four-momentum of the proton $(E_p, \vec{p}_p)$ and transfer four-momentum
defined by the electron arm $(\omega, \vec{q})$ it is possible to calculate the missing mass of the
unobserved $^{11}$B nucleus $m_{miss}$ and extract its excitation spectrum $E_{x}$:
\begin{equation} \label{eq:eq4}
E_{x}= m_{miss} -m_{^{11}B} = \sqrt{(\omega + m_{^{12}C} - E_p)^2 - (\vec{q} - \vec{p}_p)^2}-m_{^{11}B}
\end{equation} 

\begin{figure}[!ht]
\begin{center}
\includegraphics[scale=0.26]{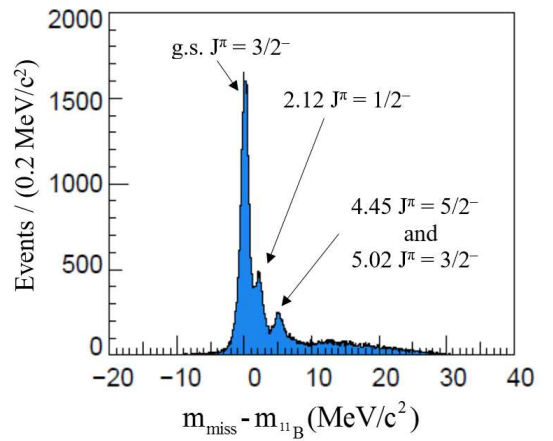} 
\caption{(colour online) The $^{11}$B excitation energy spectrum of the $^{12}C(e,e'p)^{11}B$ reaction (Setting 2).
The peak centered at 0 MeV/c$^2$ belongs to the ground state and has a width of 1.41 MeV/c$^2$ FWHM. 
The second peak is the first excited state at 2.12 MeV and the third peak is caused by higher excited 
states at 4.45 MeV and 5.02 MeV \cite{ajzenberg_68}.} 
\label{Fig7}
\end{center}
\end{figure}

Fig. \ref{Fig7} shows the excitation spectrum of the $^{11}$B nucleus for kinematical setting 2 from 
table \ref{Tab3}. The spectrum is dominated by the ground state having an experimental width 
of 1.41 MeV/c$^2$ FWHM.

\subsection {Momentum resolution}

In order to estimate the momentum resolution of the SOS, the following assumptions were made. For 
setting 2, protons (265 MeV/c) have a kinetic energy of 37 MeV and according to the Bethe-Bloch equation 
\cite{beringer_12} they lose 0.43 MeV of energy for a 30 mg/cm$^2$ $^{12}$C target transit. Since 
protons created deeper inside the target will have less energy loss, the value of 0.43 MeV
will be used as an upper limit for the estimate of the proton's energy uncertainty $\Delta E_p$. This
can also be expressed as the proton's momentum uncertainty $\Delta |\vec{p_p}|$:
\begin{equation} \label{eq:eq5}
\Delta |\vec{p_p}| = \frac{E_p}{|\vec{p_p}|}\Delta E_p \simeq 3.676  \; \Delta E_p 
\end{equation} 
The average energy loss of the electron beam (E = 180 MeV) inside the $^{12}$C target is up to 
57.6 keV and up to 58.4 keV for scattered electrons (E' $\simeq$ 115 MeV). In order to simplify 
calculations, the electron's momentum uncertainty is assumed to come exclusively from the beam's energy
loss ($\Delta E \simeq$ 58 keV).
Using equation (\ref{eq:eq4}) the uncertainty of the excitation spectrum $\Delta E_x$ can
be easily calculated:
\begin{align} \label{eq:delta_e_x}
\begin{split} 
(\Delta E_x)^2 &= \Bigg(\frac{\partial E_x}{\partial E} \Delta E\Bigg)^2 + 
\Bigg(\frac{\partial E_x}{\partial E'} \Delta E'\Bigg)^2 + \Bigg(\frac{\partial E_x}{\partial |\vec{p_p}|} 
\Delta |\vec{p_p}|\Bigg)^2   \\
  &\simeq 1.02 \; (\Delta E)^2 + 1.02 \; (\Delta E')^2 + 0.0795 \; (\Delta |\vec{p_p}|)^2 \\
&\simeq 1.02 \; (\Delta E)^2 + 1.02 \; (\Delta E')^2 + 1.074 \; (\Delta E_p)^2
\end{split}
\end{align}
The last line in equation \ref{eq:delta_e_x} was obtained by using equation \ref{eq:eq5}.
Finally, an error-less reconstruction of the proton momentum in spectrometer A is assumed. For
$\Delta E_x$ = 1.41 MeV, $\Delta E_p$ = 0.43 MeV and $\Delta E$ = 0.058 MeV the upper limit of 
SOS's momentum resolution is estimated as $\Delta E' \lesssim$ 1.31 MeV.

Expressed in terms of relative momentum resolution this amounts to 1.31/115 $\simeq$ 1.14\%. This 
relatively large value results from the missing momentum resolution of the participating proton.
How much multiple scattering of low energy protons influences the momentum resolution 
of spectrometer A was not investigated in the framework of this work, since it was not essential
for the creation of the transfer matrices. 

If the full momentum acceptance range of the SOS is reduced, the corresponding width of the ground
state peak is reduced to 1.2 MeV/c$^2$ FWHM. The same effect is also observed if one reduces the
momentum acceptance range of spectrometer A. This is to be expected, since the acceptance ranges 
of both spectrometers are correlated with each other via the quasielastic proton knockout process. 
Consequently, it looks like momentum determination in a central region works 
better than towards the edges of the momentum acceptance. Or in other words, the magnet optics are
not free of aberration.

\subsection {Angular resolution at target}

The angular resolution at target was estimated using the procedure described in \cite{korn_94}. 
Individual holes of the sieve collimator are described with the aperture angle $\theta_{\epsilon}$. For 
a distance of 66 cm from target to collimator, the aperture angle of the collimator's 2 mm and 3 mm holes are 3.03 mrad and 4.55 mrad, respectively.

If the distributions of dispersive angle $\theta_0$ and non-dispersive angle $\phi_0$ 
have corresponding widths $\Delta \theta_0$ and $\Delta \phi_0$, the following upper estimations of
angle resolutions can be made:
\begin{align} \label{eq:res_ang}
\begin{split} 
\sigma_{\theta_0} &\leq \sqrt{(\Delta \theta_0)^2-\theta_{\epsilon}^2} \\
\sigma_{\phi_0} &\leq \sqrt{(\Delta \phi_0)^2-\theta_{\epsilon}^2}
\end{split}
\end{align} 
The data for estimation of angular resolution was selected by the cut around the nominal 
hole positions as follows:
\begin{equation} \label{eq:ang_cut}
(\theta_0 - \theta_0^{nom})^2 + 4 \cdot (\phi_0 - \phi_0^{nom})^2 < (0.6^{\circ})^2
\end{equation}
Relative to the hole center used cuts were $\pm$0.6$^{\circ}$ for $\theta$ and $\pm$0.3$^{\circ}$ 
for $\phi$. The FWHM values of the angle distributions, averaged over all kinematical settings and number of 
holes, were $\Delta \theta_0 = $ 11.6 mrad (3 mm holes) and 11.0 mrad (2 mm holes). Finally, 
after averaging over hole sizes a resolution value of $\sigma_{\theta_0} =$ 10.6 mrad was 
obtained for the dispersive angle at the target. 

For the non-dispersive angle the averaged FWHM value of $\Delta \phi_0 = $ 3.3 mrad was only 
determined for holes having 2 mm diameter (equation (\ref{eq:res_ang}) could not be used 
for 3 mm holes, since the uncertainty was smaller than the aperture angle). A resolution 
value of $\sigma_{\phi_0} =$ 1.3 mrad was obtained for the non-dispersive angle at the target.

\begin{figure}[!ht]
\begin{center}
\includegraphics[scale=0.24]{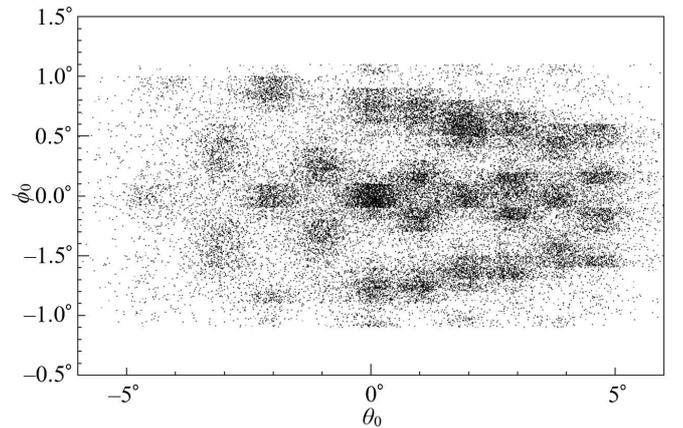}
\caption{Calculated target coordinates $\theta_0$ and $\phi_0$ for the electron momentum 
of 125 MeV/c and cuts on coincidence peak $| T_{A-SOS}| < $ 2 ns, on $^{11}$B ground state 
$|E_x| <$ 2.5 MeV/c$^2$, on wire multiplicity $>$ 5 in both chambers and on SOS momentum 
acceptance -13\% $< \delta p_0 <$ 16\%. Measurement was performed with sieve collimator.}
\label{Fig8}
\end{center}
\end{figure}

RAYTRACE \cite{kowalsky_86, kowalsky_87} simulations of particle trajectories having
$\theta_0 >$ 0 predict a smaller value of $\sigma_{\phi_0}$ compared to those having $\theta_0 <$ 0 
\cite{baumann_dipl}. This can be clearly observed in Fig. \ref{Fig8}: with increasing 
of $\theta_0$, distributions of the individual holes get narrower. Described effect can be
explained that the particles having positive $\theta_0$ (going down inside the SOS) are slightly 
focused in the non-dispersive direction by the magnet's fringe fields and, therefore, angular
acceptance has a trapezoidal shape.

\subsection {Experimental determination of transfer matrix}

The full coverage of aperture edges of the two above mentioned collimators is achieved 
by placing the sieve collimator at two different positions. In those two positions none of 
the holes overlap with each other, so that the number of coordinates used in determination of the
transfer matrix is doubled. 

\begin{figure}[!ht]
\begin{center}
\includegraphics[scale=0.21]{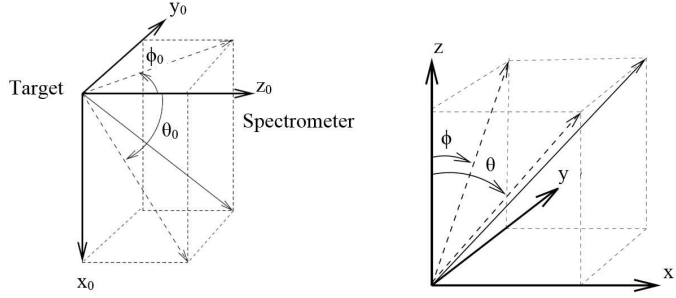}
\caption{Definition of target coordinate system (left) and chamber coordinate system 
(right), from \cite{korn_94}.}
\label{Fig9}
\end{center}
\end{figure}

In the target coordinate system (Fig. \ref{Fig9}) positions of the holes are defined
by dispersive angle $\theta_0$ and non-dispersive angle $\phi_0$. Additionally, in 
the dispersive xz-plane we also have the difference from the central momentum 
$\delta p_0 = \Delta p/p$ and in the non-dispersive yz-plane we have the vertex coordinate 
$y_0$ in the direction of the beam. By using drift chamber measurements, the position of
each hole can be determined and mapped in terms of the chamber coordinates 
$(\theta_{ch}, x_{ch}, \phi_{ch}, y_{ch})$. Now it is possible to write the dispersive 
target coordinates $\theta_0$ and $\delta p_0$ as functions of the dispersive chamber 
coordinates $(\theta_{ch}, x_{ch})$, as well as the non-dispersive target coordinates 
$\phi_{0}$ and $y_{0}$ as functions of $(\phi_{ch}, y_{ch})$. The coefficients of 
these functions form the transfer matrix. In addition, we have the path length 
$l$, which is needed for correction of coincidence time.

An example of transfer matrix coefficients and corresponding functions obtained for 
the kinematical setting 1 and cuts on coincidence peak $| T_{A-SOS}| < $ 2 ns, on the
$^{11}$B ground state $|E_x| <$ 2.5 MeV/c$^2$, on wire multiplicity $>$ 5 in X- 
and Y-chamber, and on SOS momentum acceptance -13\% $< \delta p_0 <$ 16\% is:
\begin{align} \label{eq:trans_mat}
\begin{split} 
\delta p_0 &= 1.85 \cdot x_{ch} - 0.018 \cdot \theta_{ch} + 0.0014 \cdot x_{ch} \cdot \theta_{ch} \\
\theta_0 &= 8.512 \cdot x_{ch} - 0.520 \cdot \theta_{ch} \\
y_0 &= 0.383 \cdot y_{ch} - 0.045 \cdot \phi_{ch} \\
\phi_0 &= 5.011 \cdot y_{ch} + 0.037 \cdot \phi_{ch} \\
l &= 163 \; cm
\end{split}
\end{align}
where $\delta p_0$ is expressed in percentage, all angles are expressed in mrad and all 
lengths in cm. Angles at target $\theta_0$ and $\phi_0$ reconstructed using this 
matrix can be seen in Fig. \ref{Fig8}.

The obtained transfer matrices are not suitable for reconstruction of the target coordinate 
$y_0$, because they have been produced by using only one carbon foil as target. In order 
to be sensitive to $y_0$, calibration of the SOS has to be performed with a stack of
carbon foils with well defined distances between the individual foils.

\section{Test measurement of the $\mathrm{p(e,e'\pi^+)n}$ reaction} \label{electroproduction}

The $\mathrm{p(e,e'\pi^+)n}$ reaction was measured at an invariant mass of $W$ = 1084.3 MeV ($\sim$5 MeV
above threshold) and a four-momentum transfer of $Q^2$ = 0.078 (GeV/c)$^2$. Other parameters of 
the measured kinematical setting are summarized in table \ref{Tab4}. Liquid hydrogen 
inside a cylindrical cell was used as target. The cell, 2 cm in diameter, was made of a 50 $\mu$m 
Havar foil as boundary. Density fluctuations inside the liquid 
hydrogen were avoided by recirculation of the liquid hydrogen and by rastering the electron 
beam in transverse directions.
\begin{table}[h!]
\caption{Central kinematical parameters for the $\mathrm{p(e,e'\pi^+)n}$ reaction: $\epsilon$ 
is the transversal polarization of the virtual photon, $E$ is the beam energy, $p_{A}$ and 
$\theta_{A}$ are the central momentum and the angle of spectrometer A, $p_{SOS}$ and 
$\theta_{SOS}$ are the central momentum and the angle of SOS.} 
\begin{center}
\begin{tabular}{cccccc} 
$\epsilon$     & $E$    & $p_{A}$   & $\theta_{A}$  &  $p_{SOS}$  & $\theta_{SOS}$  \\
            & (MeV)  & (MeV/$c$) & ($^{\circ}$)  &    (MeV/$c$)      &  ($^{\circ}$) \\ \hline 
0.902   & 855    & 656.2 &    22.5       &    82.7          & 44.5 \\ 
\end{tabular} 
\end{center}
\label{Tab4} 
\end{table} 

\begin{figure}[!ht]
\begin{center}
\includegraphics[scale=0.38]{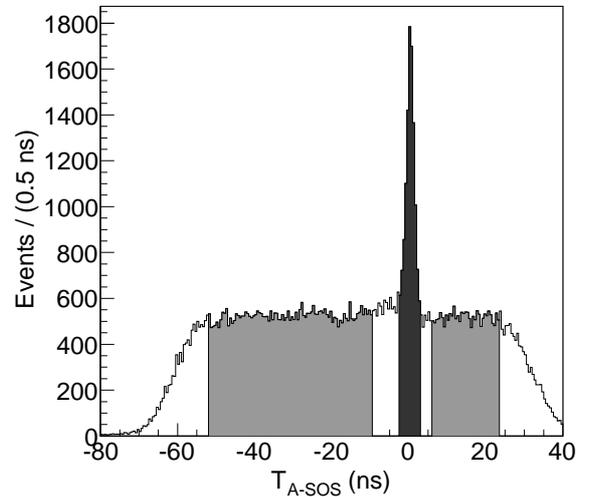} 
\caption{\label{Fig10} Coincidence time distribution. The dark gray area includes
true coincidences and the light gray area contains only random coincidence events used 
for the background estimation.}
\end{center}
\end{figure}

The scattered electron in spectrometer A was detected in coincidence with the produced 
charged pion in the SOS. After the correction of the coincidence time for the particle 
path length in both spectrometers and adjustment of delays in electronics, a coincidence 
peak having a time resolution of 2.46 ns FWHM was obtained, see Fig. \ref{Fig10}. The 
true electron-pion events were selected using a cut $-$2 ns $\le T_{A-SOS} \le$ 2.5 ns.
The contribution from random coincidence was estimated using data from the two 
sidebands: $-$52 ns $\le T_{A-SOS} \le$ $-$10 ns and 6 ns $\le T_{A-SOS} \le$ 23 ns.

\begin{figure}[!hb]
\begin{center}
\includegraphics[scale=0.38]{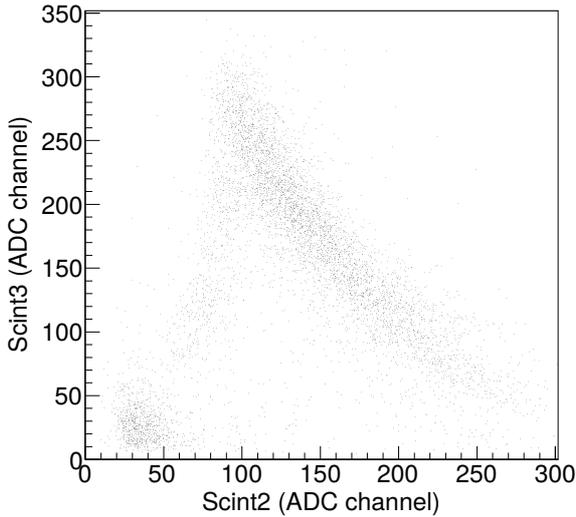} 
\caption{\label{Fig11} 2D distribution of deposited energy in the second vs. the third scintillator, obtained by using the cut on the coincidence peak and the fifth scintillator as a veto detector. Minimum ionizng particles are in the lower left corner of the figure and they are clearly separated from remainig particles. The rising and the falling slope correspond to pions which are stopped in the fourth and the third scintillator, respectively.}
\end{center}
\end{figure} 

Further reduction of background from random coincidences was achieved using 
additional cuts. Hence, a cut on the reconstructed electron momentum and a cut on the 
electron vertex reconstructed in spectrometer A were imposed. In case of the SOS, cuts 
were imposed on dispersive target coordinates and on energy deposited in the 
second and the third scintillator, which were used to suppress the background from 
minimum ionizing particles, see Fig. \ref{Fig11}.

\begin{figure}[!ht]
\begin{center}
\includegraphics[scale=0.56]{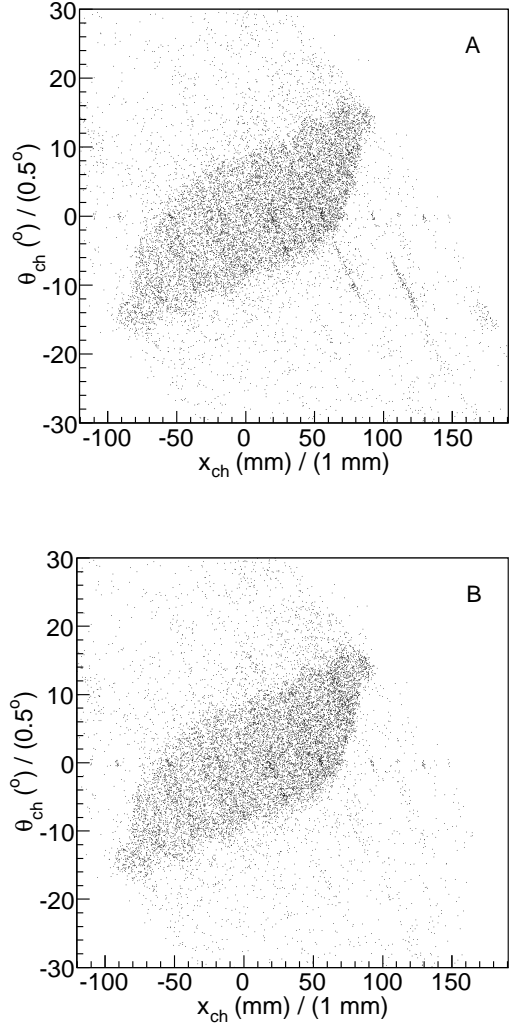} 
\caption{2D distribution of dispersive SOS drift chamber coordinates $x_{ch}$ vs. 
$\theta_{ch}$. Figure A shows the distribution without left-right correction: 
the missing data can be clearly seen as empty stripes on the right hand side of central 
distribution. Since the angle of the missing data is reconstructed with a wrong sign, 
those data appear as stripes in the lower right corner of figure A. Figure 
B shows the data with left-right correction included: central distribution does 
not contain any empty stripes.}
\label{Fig12}
\end{center}
\end{figure}

As already observed in \cite{ding_phd}, when a particle passes very close to signal 
wires, the left-right decision can turn out wrong. Instead of the real trajectory, a 
trajectory having a wrong angle sign is reconstructed. This effect can easily be seen 
by plotting the dispersive SOS drift chamber coordinates $x_{ch}$ vs. $\theta_{ch}$, 
as shown in Fig. \ref{Fig12}A. In order to correct for this issue the data stripes 
in lower right part of the distribution were isolated and their angle sign was 
changed intentionally \cite{friscic_phd}. Using this correction, the vacancies in the 
central distribution disappear (see Fig. \ref{Fig12}B). Concurrently, the number 
of true events is raised by approximately 2\%.

The effect of a wrong left-right decision is not unmistakably visible on the left 
hand side of the central distribution in Fig. \ref{Fig12} (both A and B). Either 
particles corresponding to the left part of the central distribution do not have 
trajectories for which the angle sign is easily reconstructed wrong, or the particles 
having a wrong angle sign are hidden in the central distribution. Nevertheless, if the 
latter would be the case the number of particles with wrong angle sign would still 
be very small.

\subsection{Pion decay correction}\label{sec:pion_decay_corr}

The positive charged pions $\pi^+$, on their path through the spectrometer, having 
a lifetime of $\tau_{\pi} = 26.033$  ns, may decay to a muon $\mu^+$ and a muon 
neutrino $\nu_{\mu}$. The branching ratio for this decay channel is $99.9877 \%$
\cite{beringer_12}. Decay can be described by:
\begin{equation} \label{eq:lpi_1}
\frac{N^{det}_{\pi}} {N^{tg}_{\pi}} = e^{-s/l_{\pi}} = \frac {1}{K_{decay}}
\end{equation}
where $ N^{det}_{\pi} $ is the number of detected pions, $ N^{tg}_{\pi} $ is the 
number of created pions at the target, $ K_{decay} $ is the pion decay correction 
factor, $ s $ is the length of the pion path and $ l_{\pi} $ is the pion decay length, 
which can be calculated in the following way:
\begin{equation} \label{eq:lpi}
 l_{\pi} = \gamma_{\pi} \tau_{\pi} \beta_{\pi} c  = \tau_{\pi} c \Bigg(\frac{p_{\pi}} {m_{\pi}c}\Bigg)
\end{equation}
\begin{figure}[!hb]
\begin{center}
\includegraphics[scale=0.4]{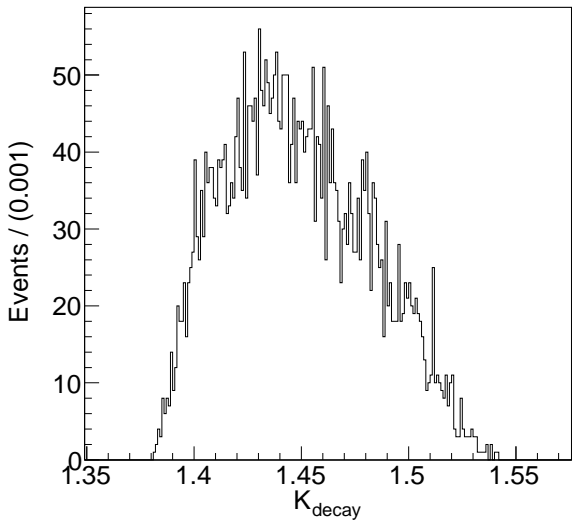} 
\caption{Distribution of the decay correction factors.}  
\label{Fig13}
\end{center}
\end{figure}

The decay correction factor $K_{decay}$ is calculated for each valid event. The calculation
of the pion decay length $l_{\pi}$ is straightforward by using equation (\ref{eq:lpi}), 
since the particle momentum $p_{\pi}$ is measured by the spectrometer. The pion path length 
$s$ is calculated using measured chamber coordinates, reconstructed target coordinates and 
the known magnetic field of the SOS's dipole. A typical distribution of the $K_{decay}$ can 
be seen in Fig. \ref{Fig13}. The data in this figure show an average value of the pion decay
correction factor of 1.45. From this we can estimate that 31\% of pions produced at the target
will decay before reaching the last detector.

\subsection{Missing mass distribution}\label{sec:mu_con}

Using the reconstructed electron and pion four-momenta, the missing mass of an unobserved neutron 
was calculated for each electron-pion pair: 
\begin{equation} \label{eq:mmiss}
m_{miss} = \sqrt{(\omega + m_p - E_{\pi})^2 -  (\vec{q} - \vec{p}_{\pi})^2}
\end{equation}
where $m_p$ is the proton mass, $\omega$ and $\vec{q}$ are energy and momentum of the virtual 
photon, $E_{\pi}$ and $\vec{p}_{\pi}$ are energy and momentum of the pion. Since it is easier 
to notice possible deviations with respect to zero, the neutron mass $m_n$ was subtracted from 
the calculated missing mass.

\begin{figure}[!ht]
\begin{center}
\includegraphics[scale=0.4]{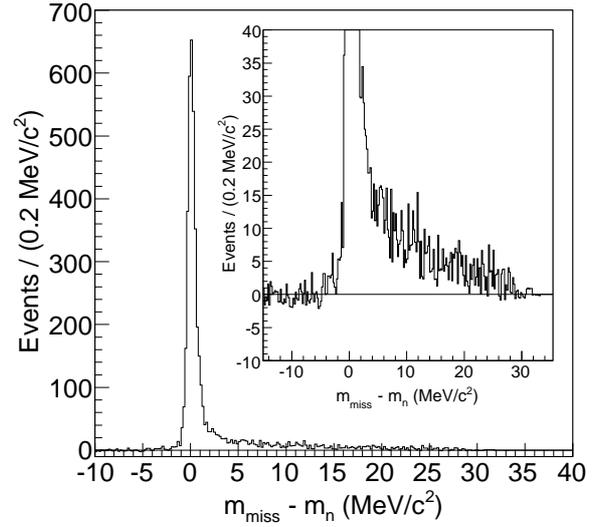}
\caption{The missing mass distribution of the background subtracted data. 
The inset shows the radiative tail region of the same missing mass
distribution.}
\label{Fig14}
\end{center}
\end{figure} 

The background subtracted missing mass distribution of the measured kinematics can be seen 
in Fig. \ref{Fig14}. As expected, the distribution shows a peak centered at bin zero, 
having an experimental width of 0.9 MeV/$c^2$ FWHM and a radiative tail on higher masses (see 
Fig. \ref{Fig14} inset).

\section{Summary and conclusion}\label{sec:sum_conc}

A new magnetic spectrometer -- SOS -- specialized for detection of low energy pions has been developed, 
built and successfully operated in electron scattering experiments at MAMI. Its detector system 
consists of volume type drift chambers and a 5-layer scintillator telescope. 

The SOS has been calibrated using the $^{12}C(e, e'p)^{11}B$ reaction. Angular resolutions at the
target were determined to be $\sigma_{\theta_0} =$ 10.6 mrad and $\sigma_{\phi_0} =$ 1.3 mrad. 
The momentum resolution was estimated to be better than 1.3\%.

The performance of the SOS to detect pions and measuring their momenta and energies was demonstrated in an
electroproduction experiment with pions produced just 5 MeV above threshold.

The SOS is fully operable and can be used in high precision measurements involving low energy pions.

\section*{Acknowledgments} \label{acknowledgments}
We would like to thank the MAMI accelerator group for the 
outstanding beam quality. This work was supported in part by the Deutsche 
Forschungsgemeinschaft with the Collaborative Research Centres 443 and 1044 
and by the Croatian Science Foundation under project HRZZ 1680.

\bibliographystyle{elsarticle-num}
\bibliography{ShortOrbitSpectrometer}

\end{document}